\documentclass[preprint]{revtex4}
\usepackage{amsmath,amssymb,graphicx}
\begin{document}

\title{Spatially resolved simulation of a radio frequency driven
micro atmospheric pressure plasma jet and its effluent}

\author{Torben Hemke, Alexander Wollny, Markus Gebhardt,
Ralf Peter Brinkmann, and Thomas Mussenbrock}
\affiliation{Department of Electrical Engineering and Information
Sciences, Ruhr University Bochum, D-44780 Bochum, Germany}

\date{\today}

\begin{abstract}
Radio frequency driven plasma jets are frequently employed as
efficient plasma sources for surface modification and other
processes at atmospheric pressure. The radio-frequency
driven micro atmospheric pressure plasma jet ($\mu$APPJ)
is a particular variant of that concept whose geometry 
allows direct optical access. In this work, the characteristics
of the $\mu$APPJ operated with a helium-oxygen mixture and its
interaction with a helium environment are studied by numerical
simulation. The density and temperature of the electrons, as well
as the concentration of all reactive species are studied both in
the jet itself and in its effluent. It is found that the effluent
is essentially free of charge carriers but contains a substantial
amount of activated oxygen (O, O$_3$ and O$_2(^1\Delta)$).\\
The simulation results are verified by comparison with experimental
data.
\end{abstract}


\maketitle

\section{Introduction}

\begin{figure}[h!]
\includegraphics[width=0.75\columnwidth]{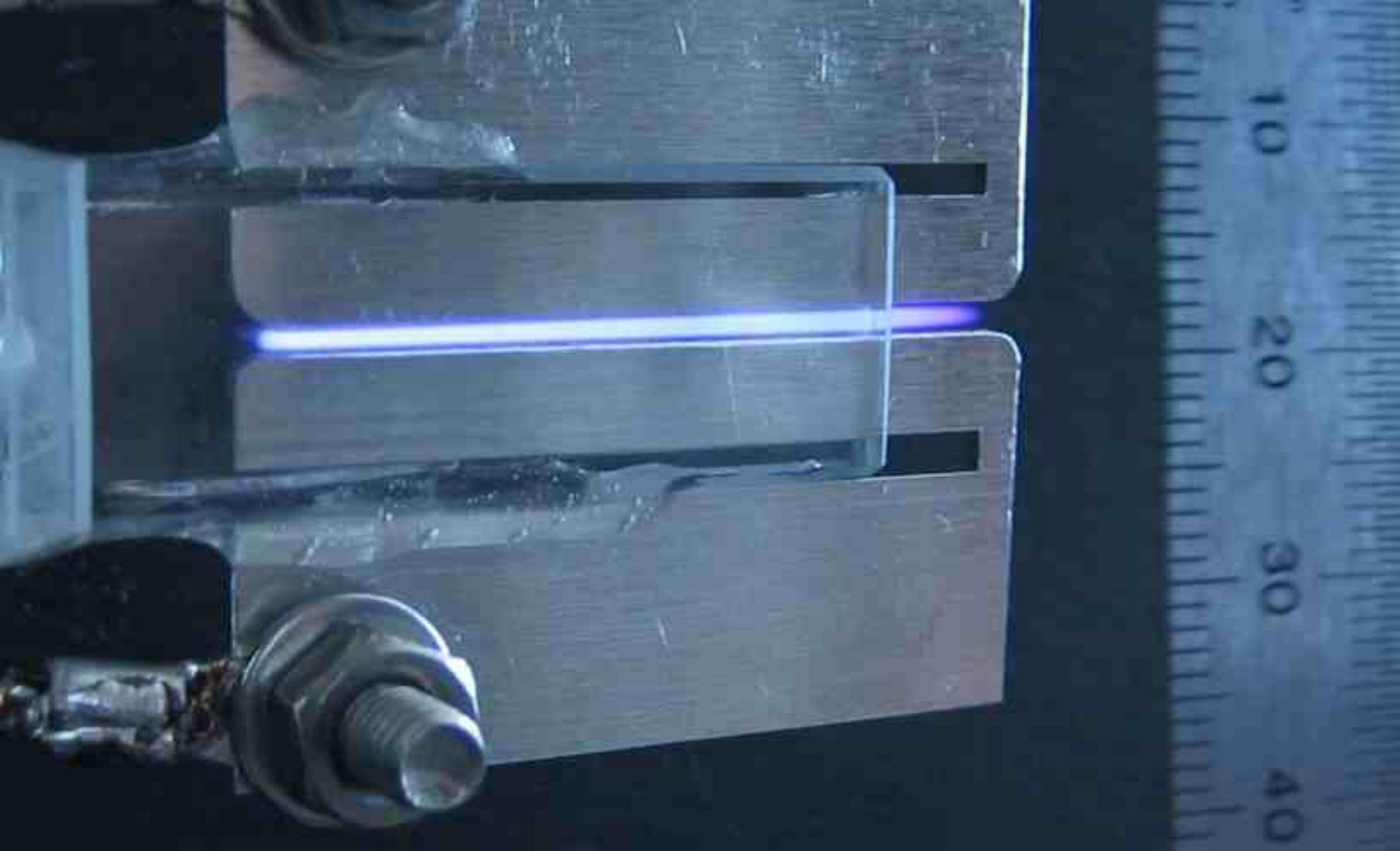}
\caption{Photograph of the micro-atmospheric pressure
plasma jet ($\mu$APPJ). (Courtesy of Dr. V. Schulz-von
der Gathen.) The discharge is confined by the top and
bottom electrode and the quartz glasses at the sides
of the jet. The stainless steel electrodes are driven at 
an RF frequency of 13.56 MHz and a nominal power of 10 W. 
The quartz cuvette allows optical access for diagnostic
purposes. The gas mixture is injected in the jet with
a high flow rate (left hand side of the jet), while
the effluent is formed outside the jet (right hand side
of the jet).}
\label{fig:jetphoto}
\end{figure}

Microplasmas operated at atmospheric pressure have recently gained
very high attention. A series of review papers and topical issues
of leading scientific journals have discussed their technological
promises as well as their scientific challenges \cite{Lee2011,
EJPD1, Morfill, BeckerBook, Becker}. A particularly popular class
of microplasmas are radio-frequency driven plasma jets, originally
proposed by Selwyn and co-workers in 1998 \cite{Selwyn}. Many groups
have investigated their prospect for surface modification; the
studied processes include etching of tungsten, deposition and
etching of silicon oxide,  and cleaning of thermolabile surfaces
from  contaminants \cite{Selwyn1998, Selwyn1999, Selwyn2001, Ichiki,
Sankaran, Raballand2008}. In order to improve the access for optical
diagnostics, Schulz-von der Gathen and co-workers modified the 
originally axis-symmetric design of the jet to a plane-parallel
geometry \cite{Schulz}. Their device, termed the micro atmospheric
pressure plasma jet ($\mu$APPJ), is depicted in Fig. \ref{fig:jetphoto}.
It is a geometrically and electrically symmetric capacitively coupled
discharge driven via an impedance matching network by RF power of
nominally 10 W at the frequency of 13.56 MHz. The plane-parallel
stainless steel electrodes are 30 mm long, 1 mm wide, and separated
by a gap of 1 -- 2 mm. To allow direct optical access, the sides are
made of quartz glass. The jet is operated with helium as the carrier gas
and an admixture of oxygen at a relatively high flow rate
of about 1 slm. The electron density is a few 10$^{10}$ cm$^{-3}$,
the ion densities are up to 10$^{11}$ cm$^{-3}$, and the radical
densities can reach several 10$^{15}$ cm$^{-3}$.

This work expands earlier one-dimensional simulation approaches
by Waskoenig et al. \cite{Waskoenig}. We study the $\mu$APPJ by
means of a two-dimensional numerical simulation. This allows us
to study not only the discharge region but also the effluent of the jet
which expands freely into the ambient atmosphere. In the effluent
the gas is characterized by the absence of charged particles and 
a concentration of reactive radicals. It was found that the
concentration of atomic oxygen decays only slowly outside the
plasma volume  and that the concentration of ozone even increases
for a few centimeter. These phenomena have been found before by experiments
\cite{Knake1, Knake2, Niermann, Ellerweg}.
(To show the reliability
of the simulation results we present a comparison with experimental
findings of Ellerweg et al. \cite{Ellerweg}.)


\section{Description of the numerical code}

The simulations of this study are carried out with the fluid dynamics
code \emph{nonPDPSIM} designed and realized by Kushner and co-workers.
For a detailed description of the code and its application to various
types of plasmas and gas discharges, see the publications \cite{BabaevaC,
Babaeva1, BabaevaA, BabaevaB} and the citations therein. Here, we just briefly
discuss the implemented equations and the underlying physics.

The code \emph{nonPDPSIM} simulates the dynamics of weakly ionized
plasmas in the regime of medium to high pressure. It takes into account
the physics and chemistry of charged particles -- electrons with mass
$m_{\rm e}$ and charge $-e$, ions with mass $m_j$ and charge $q_j$ --
and of the excited as well as the ground state neutrals (mass $m_j$). 
For all species $j$, the continuity equations (particle balances)
are simultaneously solved, where $\vec{\Gamma}_j$ is the particle
flux density and $S_j$ the source and loss term, respectively: 
\begin{gather}
\label{eq:j_continuity}
\frac{\partial n_j}{\partial t} = -\nabla\!\cdot\!\vec{\Gamma}_j + S_j.
\end{gather}
The fluxes are calculated from the momentum balances in the
drift-diffusion approximation evaluated in the local center-of-mass
system. $D_j$ and $\mu_j$ are the diffusion constant and the mobility
(if applicable) of the species $j$, $\vec E$ is the electrical field,
and $\vec v$ is the mass-averaged advective velocity of the medium: 
\begin{gather}
\label{eq:j_flux}
\vec \Gamma_j = n_j \vec{v} - D_j \nabla n_j + \frac{q_j}{|q_j|}
\mu_j n_j \vec E.
\end{gather}
For the electron fluid, additionally an energy balance equation is
solved which takes into account Ohmic heating and the energy losses
due to elastic and inelastic interaction with the neutrals and ions
as well as heat conduction, 
\begin{equation}
\label{eq:e_energy}
\frac{\partial}{\partial t} \left(\frac{3}{2}n_{\rm e} T_{\rm e}\right)
= \vec{j}\! \cdot\! \vec{E} - \nabla \cdot \left( -\kappa_{\rm e}
\nabla T_{\rm e} + \frac{5}{2} T_{\rm e} \vec{\Gamma}_{\rm e} \right) 
- n_{\rm e} \sum_i \Delta \epsilon_i k_i n_i.
\end{equation}
To capture the non-Maxwellian behavior of the electrons, all
electronic transport coefficients (the mobility $\mu_{\rm e}$,
the diffusion constant $D_{\rm e}$, and the thermal conductivity
$\kappa_{\rm e}$) as well as the electronic rate coefficients in
eqs. (\ref{eq:j_continuity}) and (\ref{eq:j_flux}) are calculated
by the local mean energy method: A zero-dimensional Boltzmann equation for
the electron energy distribution $f_{\rm}(\epsilon)$ and the transport
and rate coefficients is solved for the locally applicable gas
composition and various values of the electrical field. The tabulated
data  are then consulted in dependence of the fluid dynamically
calculated electron temperature $T_{\rm e}$. 

The plasma equations are coupled to a modified version of the
compressible Navier-Stokes equations which solve for the gas density
$\rho$, the mean velocity $\vec v$, and the gas temperature $T$.
Here, $p$ is the scalar pressure, given by the ideal gas law, and
$\overline{\overline\tau}$ the viscous stress tensor, related to
the velocity shear via the viscosity. The contributions to the
energy equation from Joule heating include only ion contributions;
the heat transfer from the electrons is included as a collisional
change in the enthalpy. The heat capacity and the thermal
conductivity of the gas are $c_{\rm p}$ and $\kappa$, respectively,
$\Delta h_j$ denotes the enthalpy change due to reaction $j$: 
\begin{eqnarray}
&&\frac{\partial \rho}{\partial t} = -\nabla\!\cdot\! \rho \vec{v},\\
&&\frac{\partial (\rho \vec{v})}{\partial t} = 
-\nabla p - \nabla\!\cdot\!\rho \vec{v} \vec{v} - \nabla \cdot
\overline{\overline \tau} +  \sum_i (q_i n_i - m_i \mu_i S_i) \vec{E},\\
&&\frac{\partial (\rho c_{\rm p} T)}{\partial t} = -\nabla \cdot 
(-\kappa \nabla T + \rho \vec{v} c_p T) - p (\nabla \!\cdot\! 
\vec{v}) (\overline{\overline \tau}\!\cdot\!\nabla \vec{v}) -
\sum_j \Delta h_j S_j +  \sum_i \vec{j}_i \cdot \vec{E}.
\end{eqnarray}

Finally, the potential $\Phi$ is calculated from Poisson's equation.
(The code works in the electrostatic approximation so that
$\vec E=-\nabla\Phi$.) The charge density on its right stems from
the charged particles in the plasma domain and from the bound charges
$\rho_s$ at the surfaces. The coefficient $\varepsilon=\varepsilon_0
\varepsilon_{\rm r}$ represents the permittivity of the medium:
\begin{gather}
-\nabla \cdot \left( \varepsilon \nabla \Phi \right) = \sum_j q_j n_j + \rho_{\rm s}.
\end{gather}
The surface charges are governed by a separate balance equation,
where $\sigma$ is the conductivity of the solid materials and the
subscript $s$ indicates evaluation on the surface:
\begin{gather}
\frac{\partial \rho_{\rm s}}{\partial t} = \left[ \sum_j q_j 
(-\nabla \cdot \vec{\Gamma}_j + S_j) - 
\nabla \cdot \left( \sigma (-\nabla \Phi) \right) \right]_{\rm s}.
\end{gather}

The dynamical equations are complemented by an appropriate set of
boundary conditions. Electrically, the walls are either powered or
grounded. With respect to the particle flow, they are either solid,
or represent inlets or outlets: The  flow is specified to a given
flux, while the outlet flow is adjusted to maintain the pressure.
Finally, it is worth mentioning that the actual implementation of
the equations poses some difficulties due to the vast differences
in the time scales of the dynamics of the plasma and the neutrals.
These difficulties are overcome by the methods of time-slicing
and subcycling.


\section{The details of the simulation case}

The described model is employed to simulate a $\mu$APPJ, similar to
the one depicted in Fig. 1. In our case, the electrodes are 30 mm
long, their separation is 1.8 mm. The simulation resolves two
Cartesian dimensions $x$ and $y$. In direction $z$, translational
invariance is assumed. (This assumption reflects a limitation of
the code. We believe that the corresponding modeling error is
not too severe; it is still possible to resolve the plasma
distribution between the electrodes and the evolution along the
direction of the gas flow. The main error of our ``infinitely wide''
model lies in the boundary conditions of the advective velocity
of the fluid at the quartz boundaries. It translates into the
coefficient of the Hagen-Poiseuille equation being off by about 
60\%, not too severe a problem as we control the total flux through
the jet, not the pressure differential between  and outlet.)

\begin{figure}[t!]
\includegraphics[width=0.75\columnwidth]{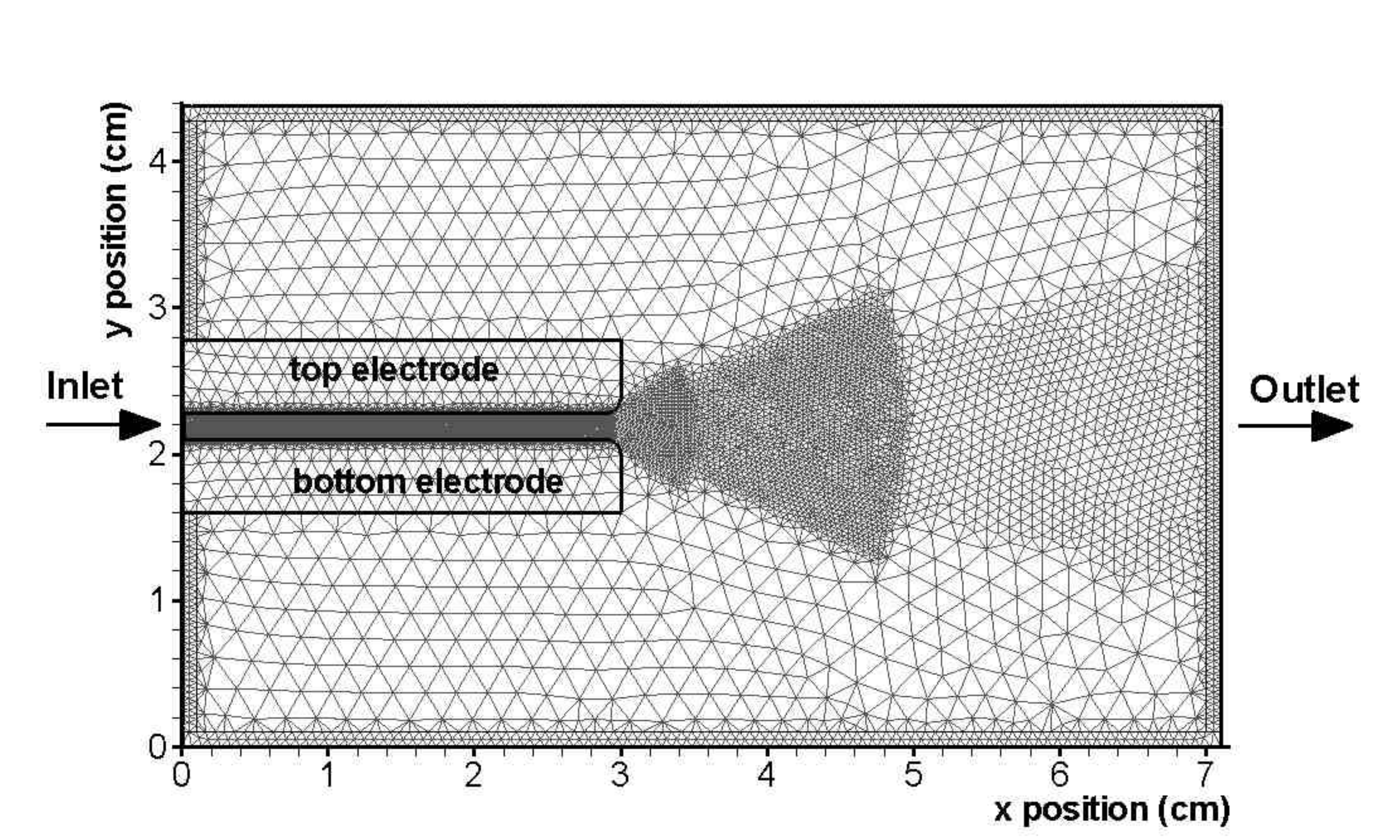}
\caption{Simulation domain with numerical mesh. The
triangular unstructured mesh consists of approximately 10000 nodes. 
Around 8000 nodes are located in the plasma region (including
the effluent), most of them between the top and bottom
electrode. The mesh becomes coarser with increasing
distance to the jet nozzle, above and under the electrodes.
On the outer border of the simulation domain we employ
the Dirichlet boundary condition $\Phi = 0$ for the
electrostatic potential (except at the electrodes).
We choose an appropriate distance between the electrodes
and the boundary to avoid parasitic discharges.}
\label{fig:simdomain}
\end{figure}

The simulation domain is depicted and explained in Fig.
\ref{fig:simdomain}. The triangular unstructured mesh consists
of approximately 10.000 nodes. 80\%  are located in the plasma
domain including the effluent, their density is highest 
between the symmetrically powered RF electrodes. The outer
boundary of the simulation domain is electrically grounded.

The device is operated as a helium-oxygen $\mu$APPJ in a pure
helium atmosphere at 10$^5$ Pa. The inlet is fed with a gas mixture
with a high flow rate of 1 slm, corresponding to a maximum advective
velocity of about 1500 cm/s. The mixing ratio of helium to oxygen
is 1000 to 5; experiments indicate that these conditions maximize
the concentration of atomic oxygen \cite{Knake1}. To maintain the
outer environment, pure helium is injected at a flow rate of 1 slm
at the left side of the simulation domain. The outlet is controlled
to maintain a constant pressure. The following species are taken
into account: Ground state neutrals O$_2$, O, O$_3$, and He,
O$_2(\nu)$, representing the first four vibrational levels of
O$_2$, the electronically excited states O$_2(^1\Delta)$,
O$_2(^1\Sigma)$, O$(^1D)$, O$(^1S)$, and He$(^2S)$, positive ions
O$_2^+$, O$^+$, and He$^+$, negative ions O$_2^-$, O$^-$, and
O$_3^-$, and electrons. The reaction chemistry applied here is described in
more detail in \cite{BabaevaA}. Following \cite{Liu}
we neglect ionized helium dimers He$_2^+$ and excimers He$_2^*$.
It has been shown that these species (which require a high electron
temperature) play a minor role when the discharge is dominated by
the reactive oxygen species; this holds for mixtures with an
O$_2$ content of $0.5\,\%$ or more. Finally, the secondary electron
emission coefficient of the surfaces is set equal to $0.1$.


\section{Simulation results}

As a result of our simulations, we obtain a complete picture of
the dynamics of the $\mu$APPJ under certain conditions.
In this manuscript we focus on 
phase-averaged quantities. The results correspond to a total
RF power deposition in the plasma of 0.5 W. (This value is
considerably lower than the nominal RF generator power of about 10 W
of the corresponding experiments. The deviation can be explained
by parasitic losses in the matching network and the feed cables.
Order-of-magnitude differences between the generator power and
the plasma heating power are also observed in other regimes
\cite{Salem, Ziegler}.) 

\begin{figure}[t!]
\includegraphics[width=0.75\columnwidth]{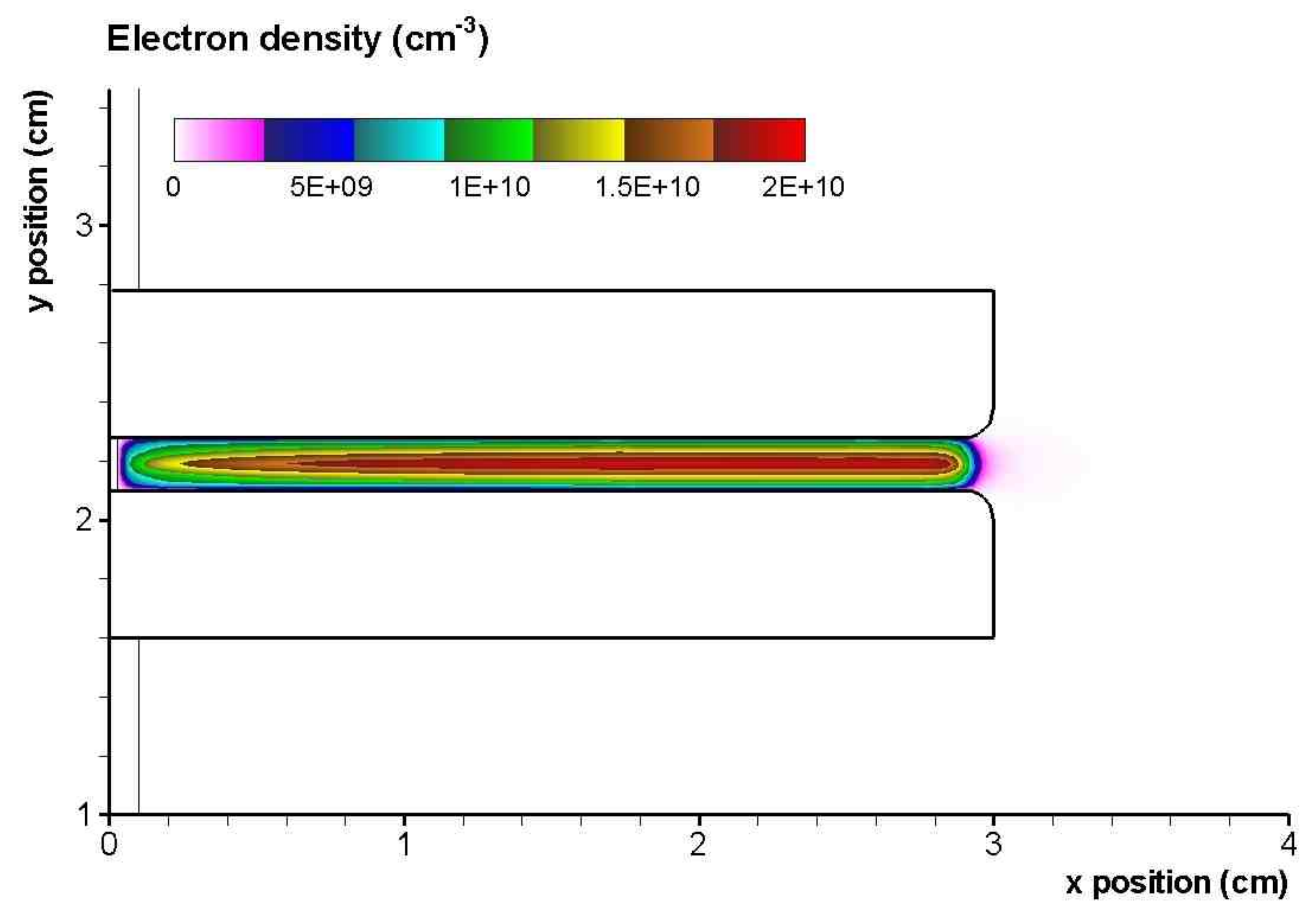}
\caption{Spatially resolved electron density $n_e$
averaged over one RF period in units of cm$^{-3}$.
The electron density is confined between the electrodes 
-- it decays rapidly in the effluent.}
\label{fig:edensity}
\end{figure}

\begin{figure}[t!]
\includegraphics[width=0.75\columnwidth]{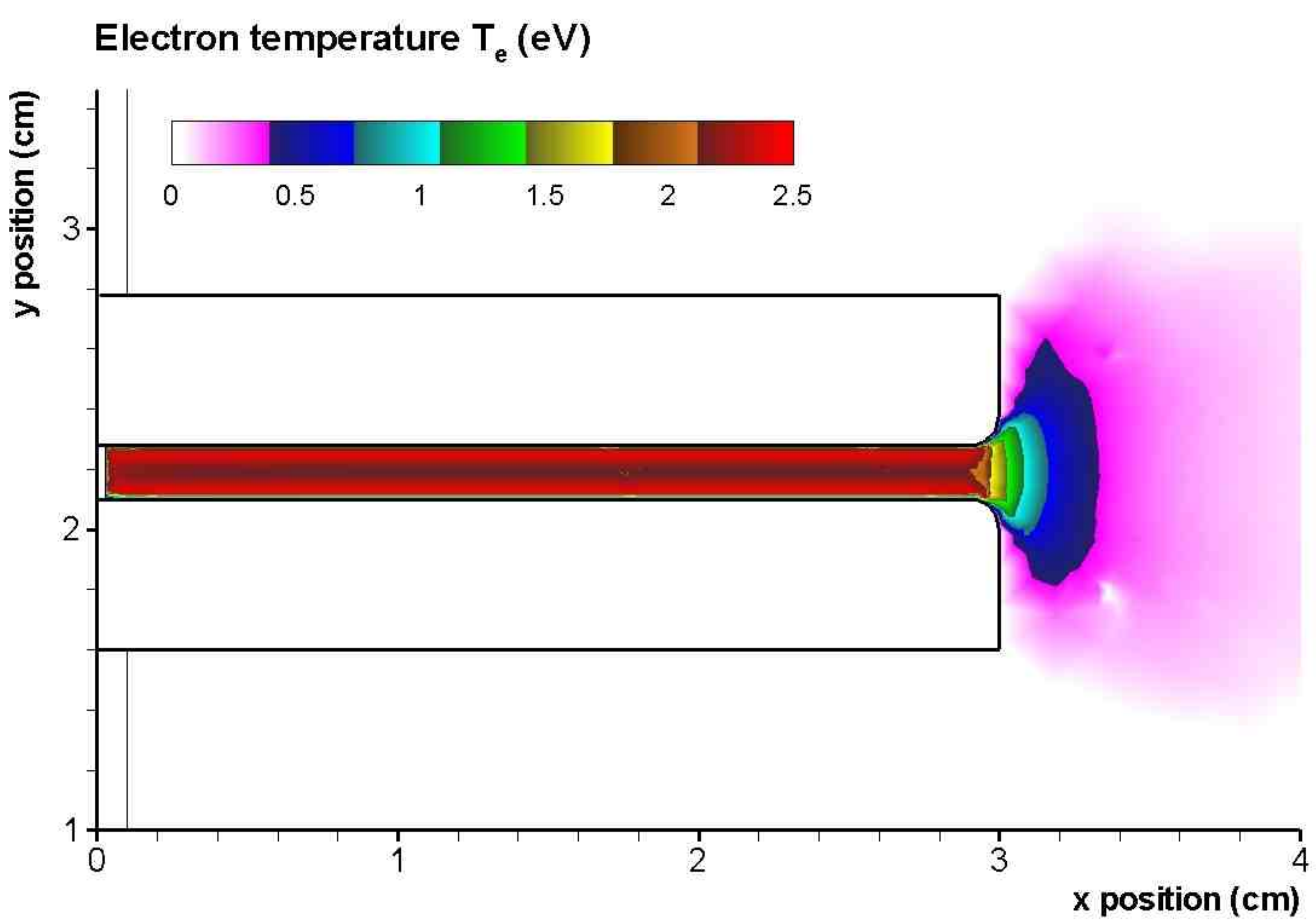}
\caption{Spatially resolved electron temperature $T_e$
averaged over one RF period in eV. The electron
temperature stays almost constant along the jet channel
while decreasing exponentially in the effluent.}
\label{fig:te}
\end{figure}

\begin{figure}[t!]
\includegraphics[width=0.75\columnwidth]{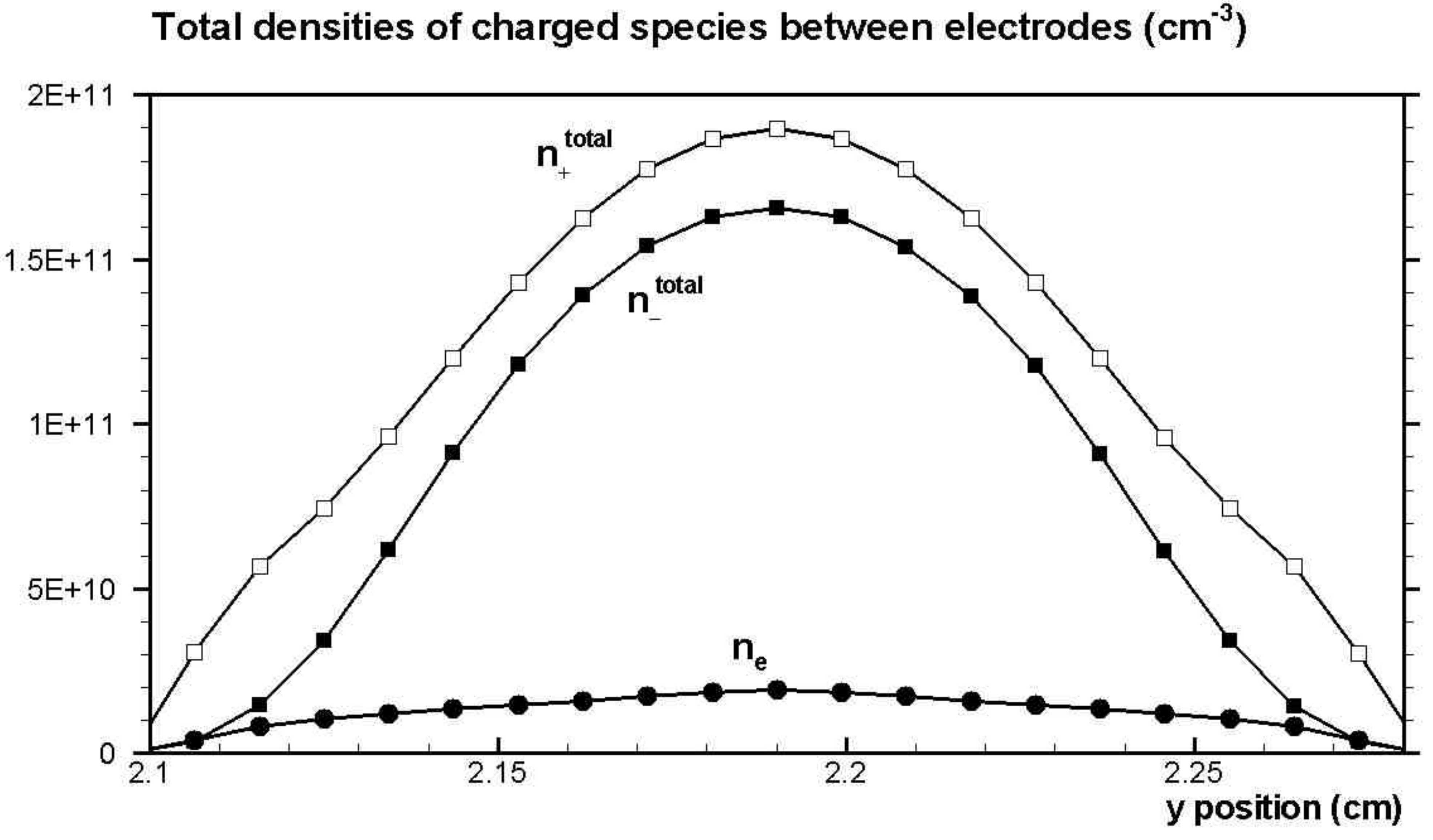}
\caption{Densities of charged species (total density of 
positive ions $n_{\text{total}}^+$, negative ions
$n_{\text{total}}^-$  and electrons $e$) averaged over
one RF period between the  electrodes across the jet
channel at $x=2$ cm in units of cm$^{-3}$. The discharge
shows a significant electronegative behavior.}
\label{fig:chargedline}
\end{figure}

\begin{figure}[t!]
\includegraphics[width=0.75\columnwidth]{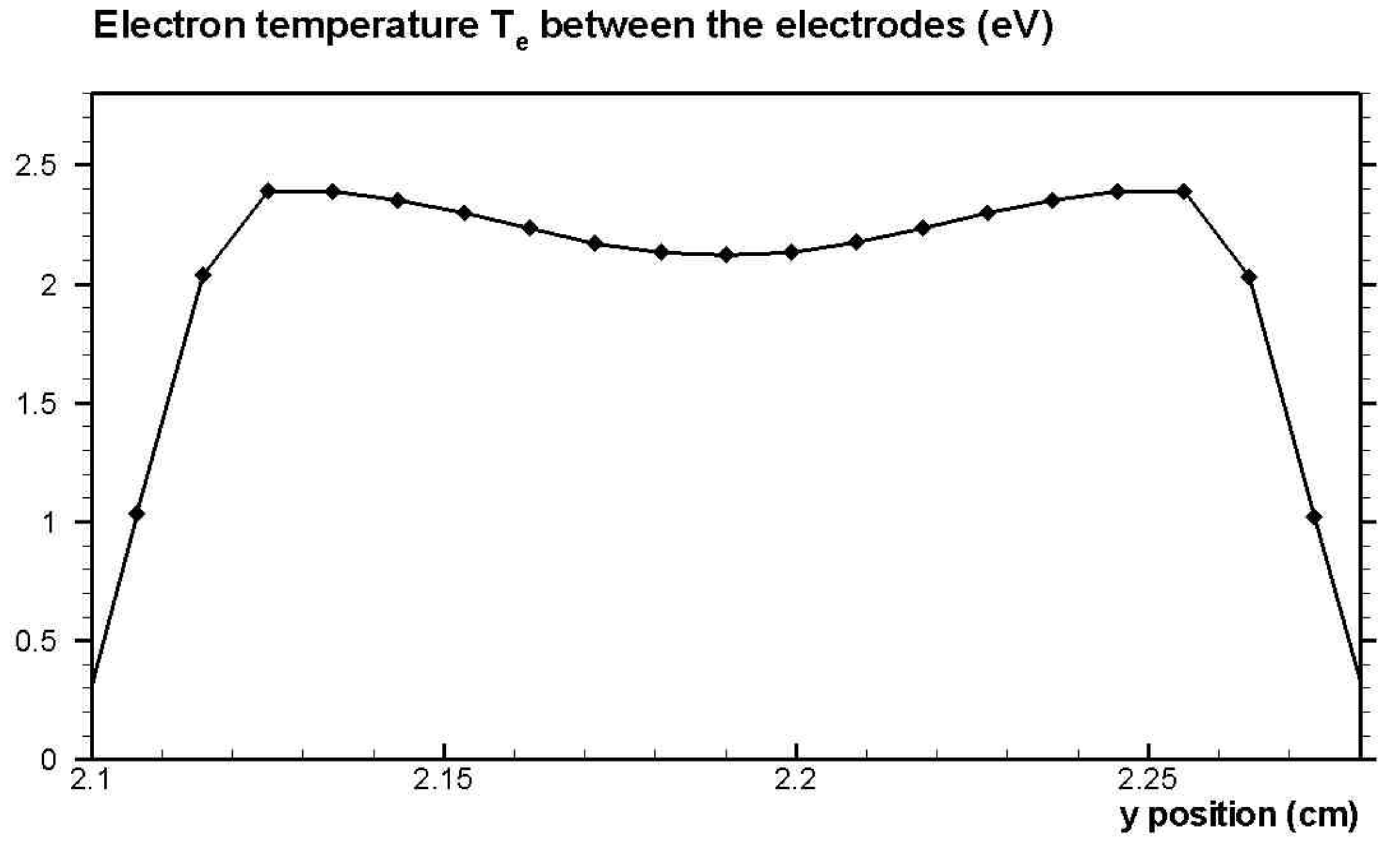}
\caption{Electron temperature $T_e$ averaged over one RF
period between the electrodes across the jet channel at
$x=2$ cm in units of eV. The maxima of the electron
temperature are located at the sheath edge.}
\label{fig:teline}
\end{figure}

Figs. \ref{fig:edensity} and \ref{fig:te} show the two-dimensional
spatial distribution of the electron density $n_{\rm e}$ and the
electron temperature $T_{\rm e}$. The 1D profiles (normal to lateral gas flow)
of the charged particle densities and the electron temperature at a position $x$=2 cm, are
shown in Figs. \ref{fig:chargedline} and \ref{fig:teline}.
Between the electrodes we notice the typical structure of a
capacitively coupled discharge. The plasma is separated in two
distinct regions, the plasma bulk and the sheath. The maximum of
the electron density is approximately 1.9$\times 10^{10}$ cm$^{-3}$. 
In the sheaths it decreases to a much smaller value. A strongly
electronegative characteristic can be observed. The charged particles 
are dominated by the oxygen species. The positive ion density is almost 
exclusively due to O$_2^+$; O$^+$ and He$^+$ are negligible. The
negative ions are O$_3^-$, O$_2^-$ and O$^-$, in descending order.
The electron temperature $T_{\rm e}$ in the plasma bulk is
approximately 2 eV, in the sheath region it is slightly larger
due to increased Ohmic heating and possibly due to secondaries.

\begin{figure}[t!]
\includegraphics[width=0.75\columnwidth]{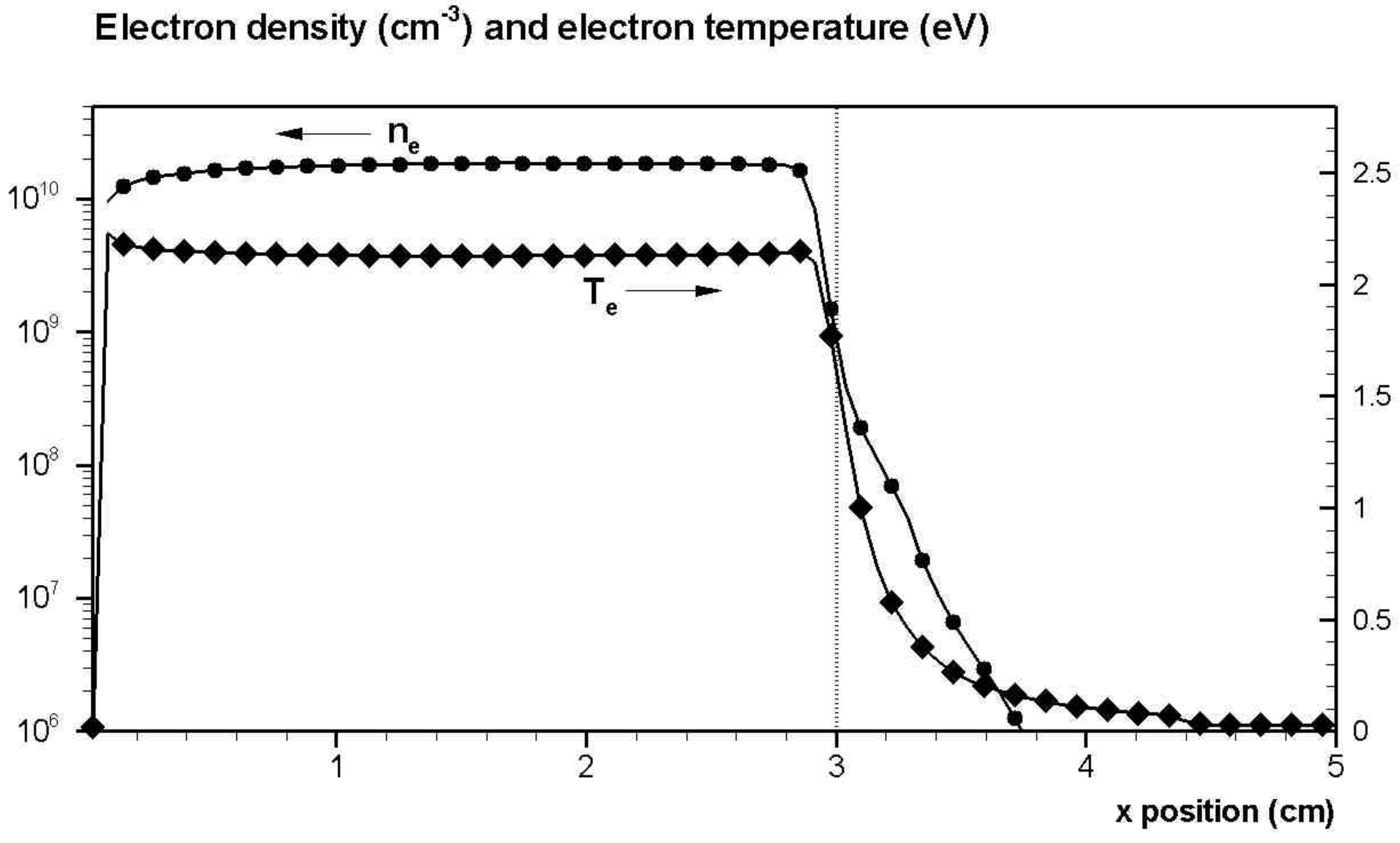}
\caption{Electron density (in cm$^{-3}$) and temperature
(in eV) in the center between the electrodes along the
jet channel, averaged over one RF period. The vertical
dotted line indicates the edge of the electrodes.}
\label{fig:nete}
\end{figure}

The effluent of the jet has the character of an afterglow. Both
the electron density and the electron temperature drop abruptly.
This can also be seen from the curves in Fig. \ref{fig:nete} 
which show the distribution of the two quantities along the
middle axis of the simulation. In the jet channel, $n_{\rm e}$
and $T_{\rm e}$ are almost constant; after the end of the electrodes
(the beginning of the effluent) 
-- indicated by the vertical dotted line -- both quantities 
decay exponentially. Fig. 8 shows that this feature
holds also for all other charged  particles except for O$_2^+$ and O$_3^-$.
We conclude from these results that the plasma dynamics
in the jet is (with respect to the $x$-coordinate) ``local''.
Generation of plasma is governed by the local electron temperature profile
which in turn is a functional of the local field distribution; 
losses are dominated by drift and diffusion to the electrodes: 
The plasma chemistry is so ``fast'' that transport due to the flow
plays a minor role and the solutions become essentially one-dimensional.
The agreement of our results with the results performed by Waskoenig et al. 
by means of a 1D-fluid model is thus not a coincidence \cite{Waskoenig}.) 

The picture changes drastically when we focus on the reactive neutral
species. (See Fig. \ref{fig:reactivespecies}.) For the neutrals the chemistry is
``slow'', and it takes a considerable time (or distance along the flow)
for the species to build up. Particularly O, O$_3$ and O$_2(^1\Delta)$
increase in the jet channel. In the effluent, their destruction is also
slow; they are transported over a distance of a few cm. The density of
ozone even increases slightly and remains constant for a certain length.
The reason for this behavior is that
the activated species are not directly generated by the plasma electrons
but by heavy particle chemistry. A particular case are the ions
O$_2^+$ and O$_3^-$. The generation of O$_2^+$ is via electron impact
ionization and therefore fast, the generation of O$_3^-$ is slow
because it involves ion chemistry. The destruction is slow for both species, 
they are transported about 1 cm into the effluent: O$_2^+$ and
O$_3^-$ form the end points of the positive and negative ion chemistry,
respectively, and their direct recombination constant is relatively small.

Further down in the effluent, therefore, only three  species play a role:
O, O$_3$ and O$_2(^1\Delta)$. Their spatial distribution is shown in
Figs. \ref{fig:spatialO}, \ref{fig:spatialO3}, and \ref{fig:spatialOmeta}.
Atomic oxygen reacts with molecular oxygen to ozone; its density thus
decays exponentially. The rate constant, however, is relatively small; 
2 cm behind the nozzle of the jet it still has a concentration of 3$\times 10^{14}$
cm$^{-3}$. The densities of ozone (2$\times 10^{15}$ cm$^{-3}$) and
excited oxygen (1$\times 10^{15}$ cm$^{-3}$) remain nearly constant
up to the outlet.

An exhaustive comparison of our simulations with experimental data 
is currently underway and will be presented in the near future. 
Nonetheless, some preliminary results may be of interest: Fig. \ref{fig:expsim} shows
the profiles of O and O$_3$ along the jet and the effluent, 
in comparison with experimental results obtained by Ellerweg et al. 
with quantitative molecular beam mass spectrometry and two-photon absorption
laser-induced fluorescence spectroscopy \cite{Ellerweg}. 
We find that the simulation results are in excellent agreement 
with the experiment. 
(Note that we changed the simulation parameters slightly to accommodate the experiment: 
The jet gap size was decreased to 1 mm. The power deposition was kept constant and the 
temperature of the electrodes – a very sensitive parameter which controls 
the production of reactive oxygen species in the effluent – was set to 330 K.)

\section{Summary and conclusion}  

In this work we employed the code \emph{nonPDPSIM} to conduct a
numerical simulation of an RF-driven atmospheric pressure plasma
jet operated with a mixture of helium and oxygen. In several aspects
our study goes beyond previous work. The two-dimensional spatial 
resolution of the simulation allows to investigate not only the
profile of the discharge parameters between the electrodes but
also their transport through the jet and into the environment.
We found a pronounced difference in the behavior of the charged
particles and the neutrals. The  charges -- $e$ and predominantly
O$_2^+$, O$^-$, O$_2^-$ -- are governed by a ``fast'' chemistry: 
In the jet, their production is  local in $x$ and their losses
are dominated by drift and diffusion to the electrodes -- this
can be understood in a 1d-picture which resolves only the $y$-axis
\cite{Waskoenig}. In the effluent, they are virtually absent.
The neutrals, in contrast, obey a ``slow'' chemistry. The
reactive oxygen species O, O$_3$ and O$_2(^1\Delta)$ build
up along the $x$-axis of the jet and are then flow-transported
far into the effluent; their distribution is genuinely
``two-dimensional''. (Actually, it will be ``three-dimensional''.
We believe, however, that the restriction of our analysis to
two dimensions does not introduce more than quantitative errors.) 

In summary, our simulation results indicate that the influence
of the jet on the environment is primarily of chemical nature.
Even in the jet, the density of the activated oxygen species
is four orders of magnitude higher than the maximum density
of the charges, and in the effluent the charges vanish completely.
This view is in agreement with the experimental findings established
by various other researchers \cite{Schulz, Waskoenig, Knake1,
Knake2, Ellerweg, Niermann, Leveille, Jeong}. In particular ozone
with a longer lifetime than atomic oxygen and the very reactive
O$_2(^1\Delta)$ are able to react with surfaces over a long
distance \cite{FridmanBook}. Excited atomic oxygen species,
in contrast, plays a relatively minor role due to its lower
concentration. 

Further work on the $\mu$APPJ will focus on the transient behavior
of the device and also on its interaction with a more complex
environment. In particular, it is planned to investigate the effect of an 
atmosphere which consists of ambient air (instead of pure helium),
and also the interaction of the effluent on a material substrate. 


\pagebreak

\section{Acknowledgment}

The authors gratefully acknowledge fruitful discussions with
Prof. M.J. Kushner and Dr. N.Y. Babaeva from the University of
Michigan at Ann Arbor. Financial support by the Deutsche
Forschungsgemeinschaft in the frame of Research Group 1123
\emph{Physics of Microplasmas} is also acknowledged.

\pagebreak


\clearpage

\clearpage

\clearpage

\clearpage

\clearpage

\clearpage

\clearpage

\clearpage

\begin{figure}[t!]
\includegraphics[width=0.75\columnwidth]{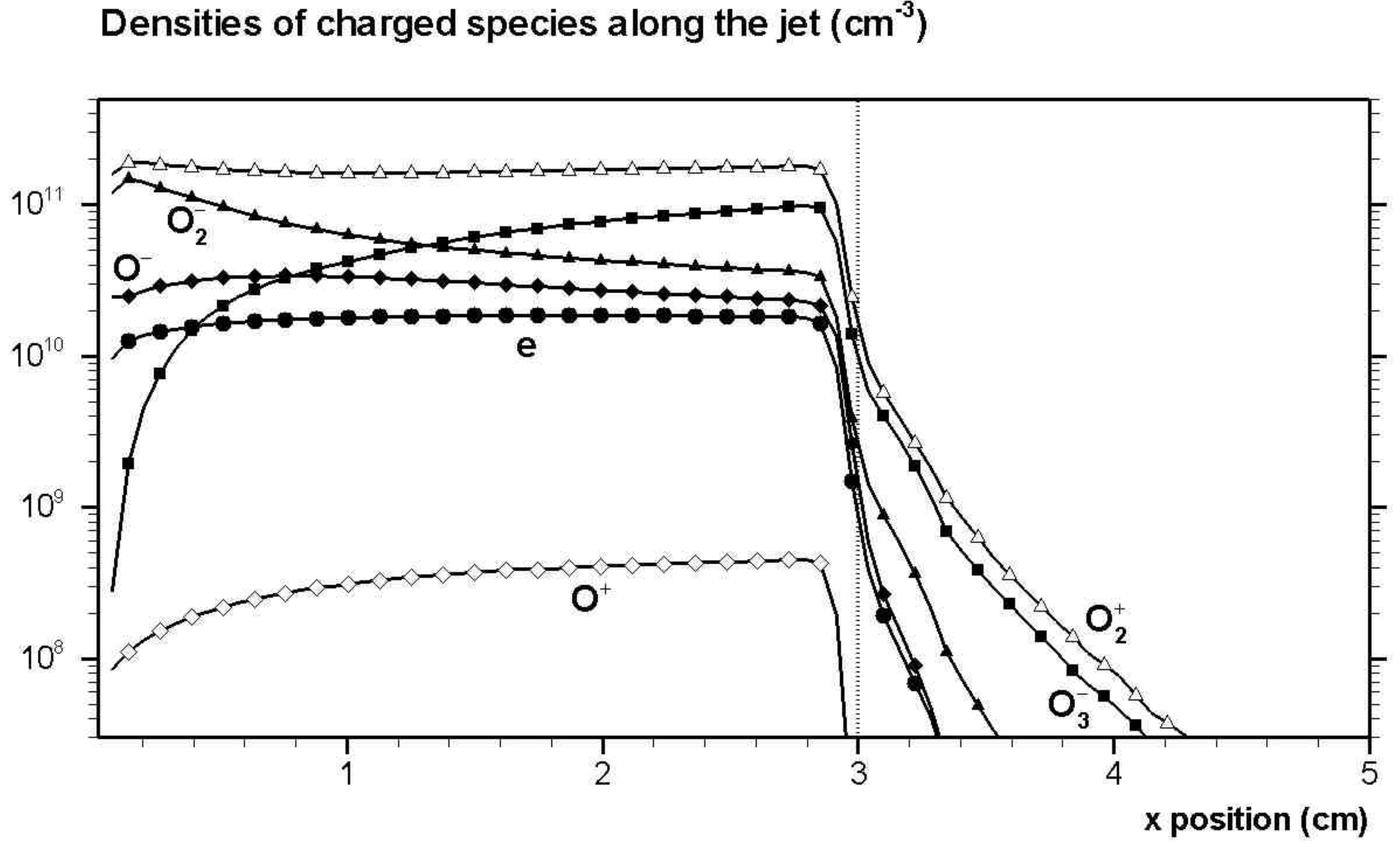}
\caption{Logarithmic-linear plot of densities of charged species
along the middle axis of the plasma jet. The dotted line
at $x=3$ cm indicates the end of the electrodes of the jet.}
\label{fig:chargedspecies}
\end{figure}

\clearpage

\begin{figure}[t!]
\includegraphics[width=0.75\columnwidth]{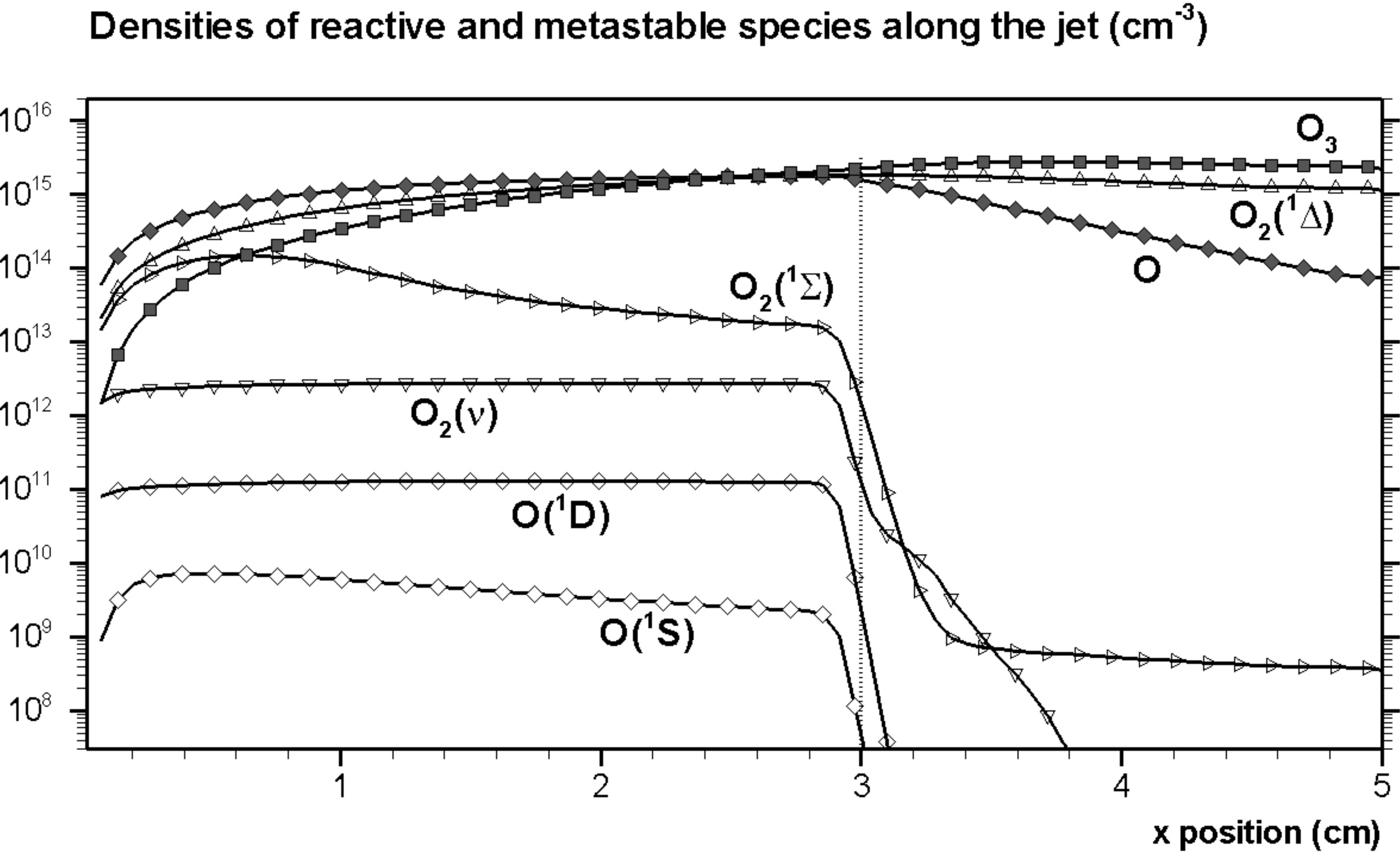}
\caption{Logarithmic-linear plot of the reactive species
and metastables densities along the middle axis of the
plasma jet. The dotted line at $x=3$ cm indicates the end
of the electrodes of the jet.}
\label{fig:reactivespecies}
\end{figure}

\clearpage

\begin{figure}[t!]
\includegraphics[width=0.75\columnwidth]{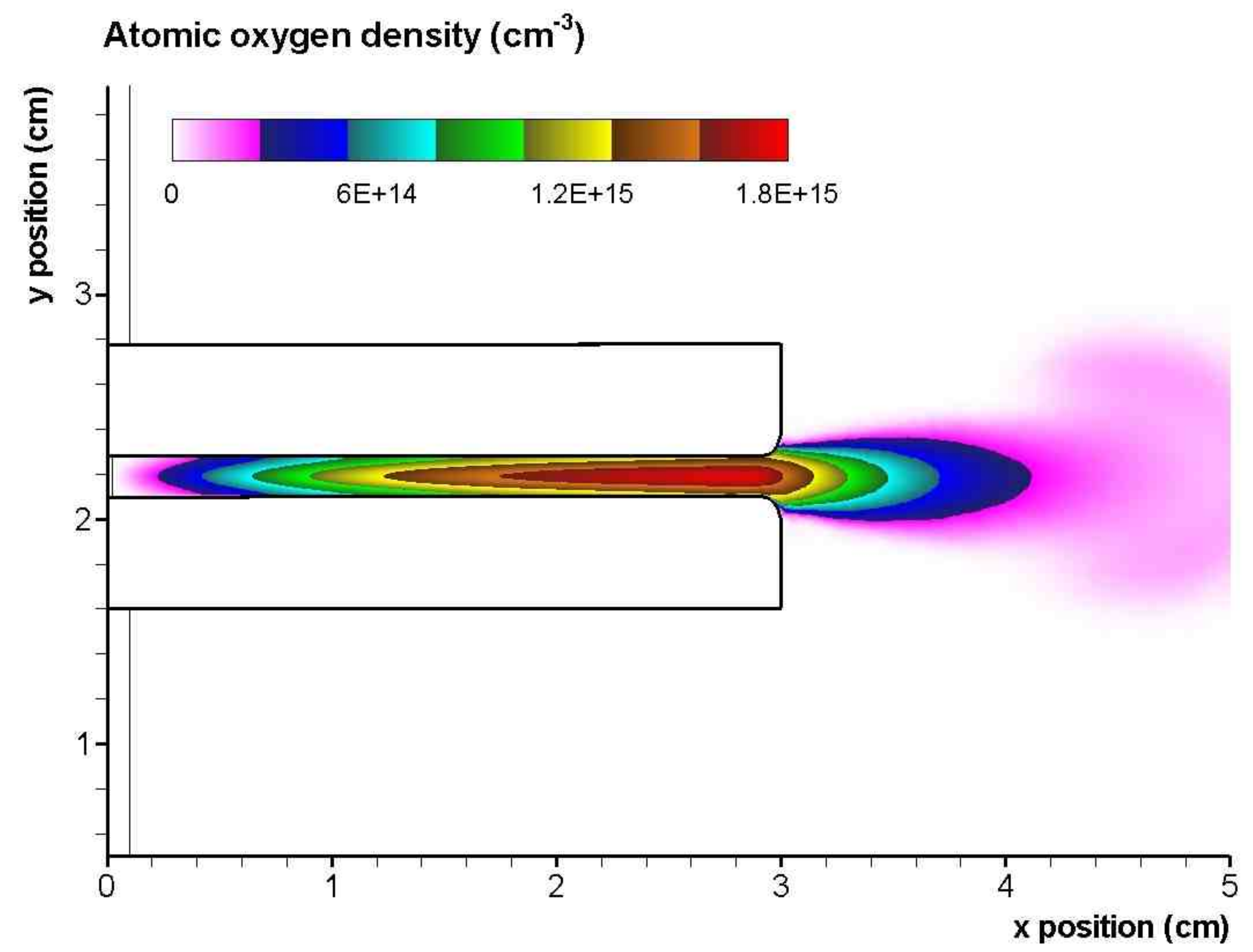}
\caption{Spatial distribution of atomic oxygen density
in units of cm$^{-3}$. After reaching the maximum of
the concentration at the nozzle of the jet the density
decays in the effluent.}
\label{fig:spatialO}
\end{figure}

\clearpage

\begin{figure}[t!]
\includegraphics[width=0.75\columnwidth]{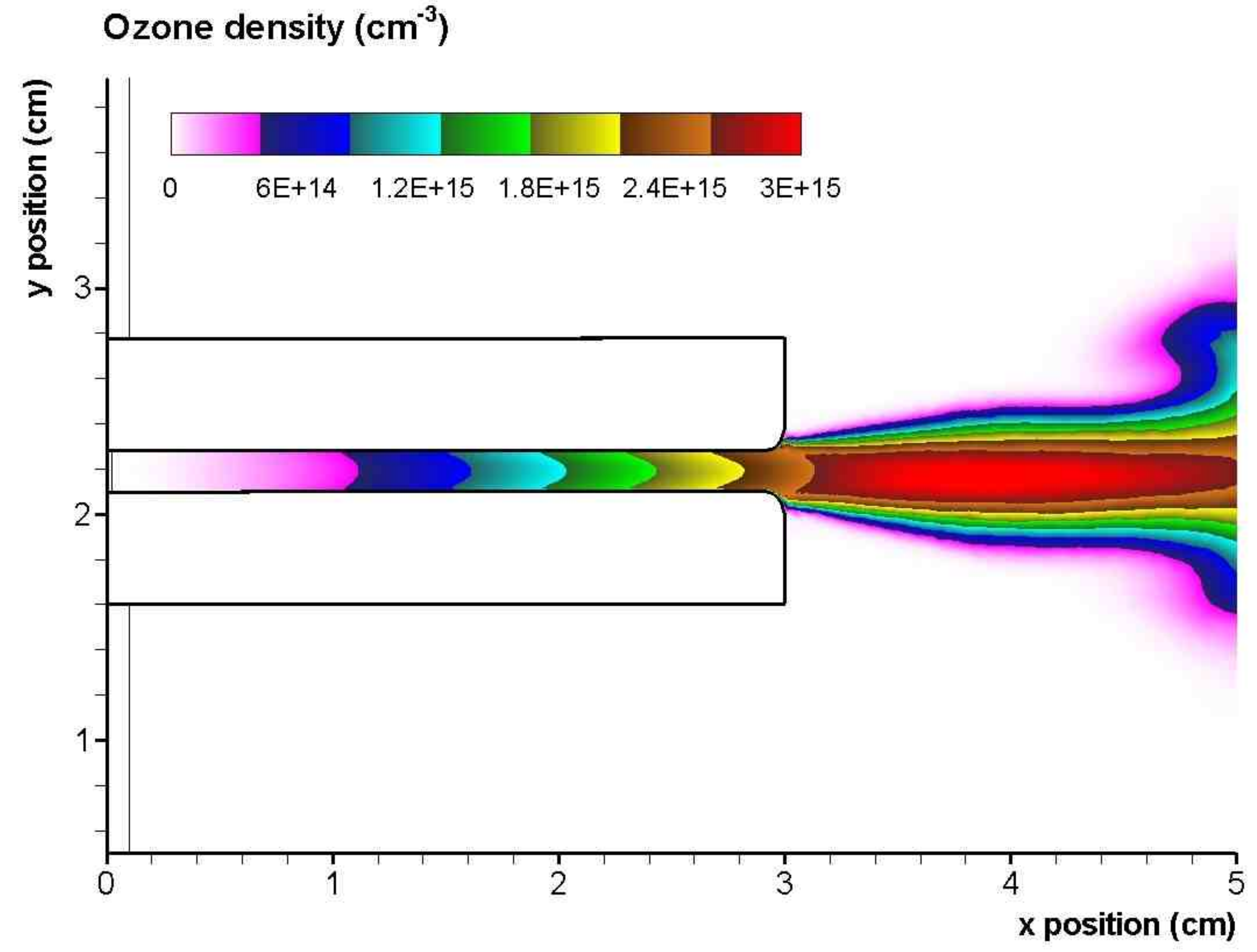}
\caption{Spatial distribution of ozone density in units
1of cm$^{-3}$. The ozone density is nearly constant across
the gap and builds up linearly along the flow. 
A nearly constant level of the ozone concentration maintains 
in the effluent.}
\label{fig:spatialO3}
\end{figure}

\clearpage

\begin{figure}[t!]
\includegraphics[width=0.75\columnwidth]{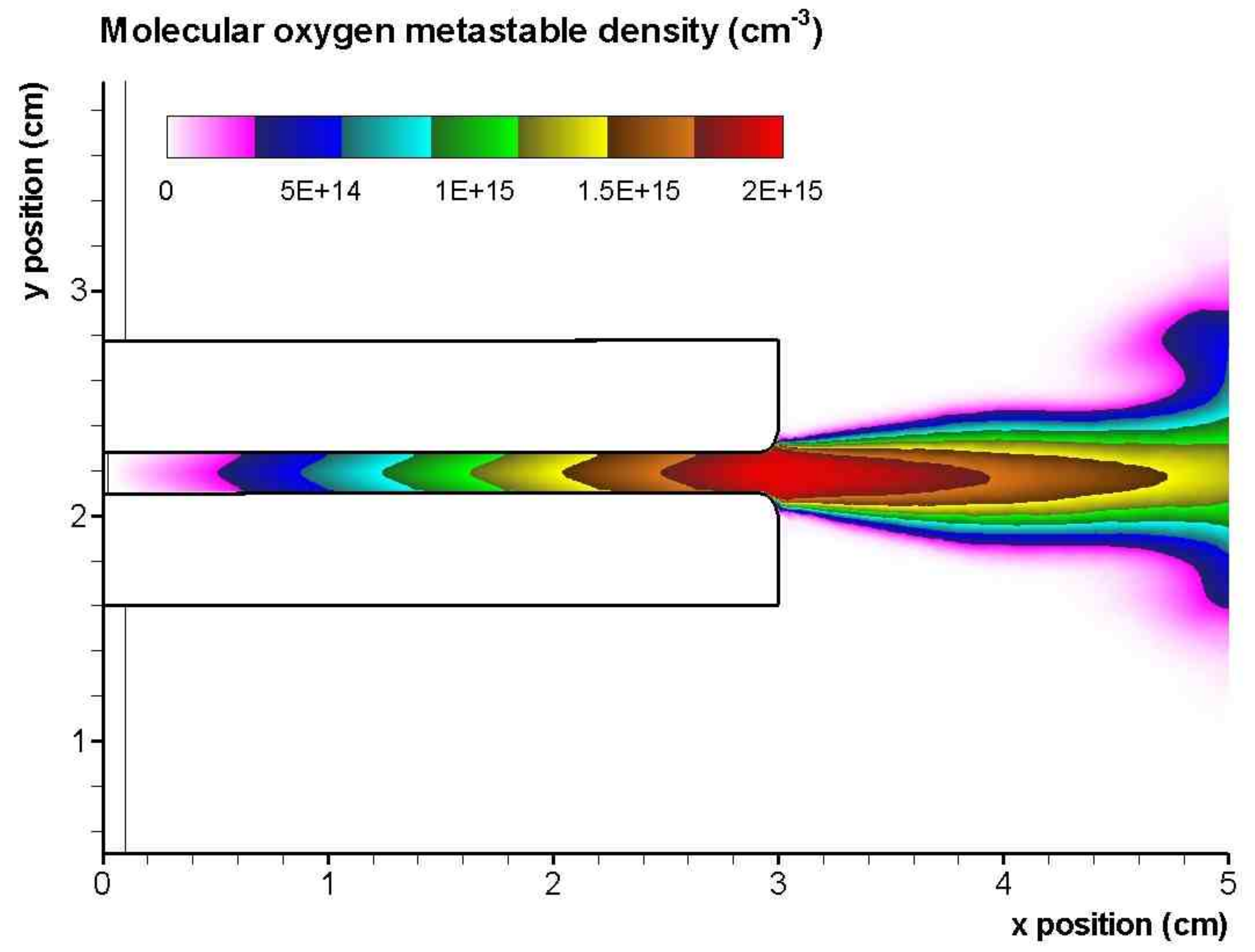}
\caption{Spatial distribution of metastable molecular
oxygen O$_2(^1\Delta)$ density in units of cm$^{-3}$.
The density of metastable oxygen is nearly constant
across the gap and grows linearly with respect to the
lateral coordinate $x$. In the effluent a slow decay
is observed.}
\label{fig:spatialOmeta}
\end{figure}

\clearpage

\begin{figure}[t!]
\includegraphics[width=0.75\columnwidth]{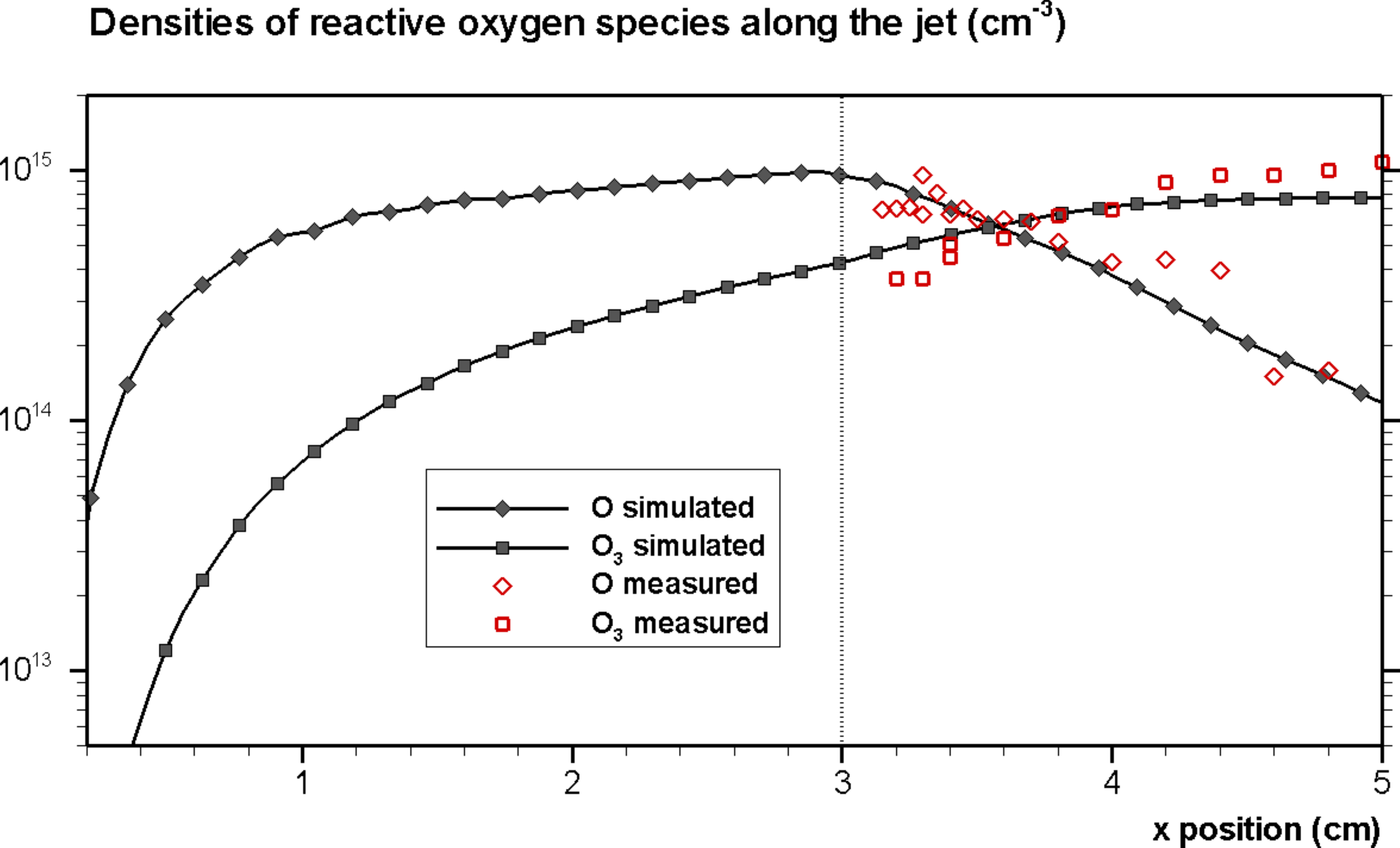}
\caption{Comparison of the spatial distribution of the densities of reactive oxygen species
in units of cm$^{-3}$ from the simulation with experimental results \cite{Ellerweg}.}
\label{fig:expsim}
\end{figure}

\end{document}